\documentclass[conference]{IEEEtran}
\ifCLASSINFOpdf
\else
\fi
\usepackage{pbox}
\usepackage{booktabs}
\usepackage{lipsum}
\usepackage{amsmath,amssymb}
\usepackage{graphicx}
\usepackage{float}
\usepackage{multirow}
\usepackage{listings}
\usepackage[hyphens]{url}
\usepackage{color}
\usepackage{colortbl,array} 
\usepackage{todonotes}
\presetkeys{todonotes}{inline}{}
\usepackage{cite}
\usepackage{tablefootnote}

\usetikzlibrary{arrows,matrix}

\usepackage[table]{}
\usepackage{array,ragged2e}
\usepackage{algorithm}
\usepackage{algpseudocode}

\newcommand\tabhead[1]{\small\textbf{#1}}

\tikzset{
  yn/.style={draw,thick,rounded corners,fill=yellow!20,inner sep=.3cm},
  bn/.style={draw,thick,rounded corners,fill=blue!05,inner sep=.3cm},
  on/.style={draw,thick,rounded corners,fill=orange!20,inner sep=.3cm},
  rn/.style={draw,thick,rounded corners,fill=red!20,inner sep=.3cm},
  greenn/.style={draw,thick,rounded corners,fill=green!20,inner sep=.3cm},
  grayn/.style={draw,thick,rounded corners,fill=gray!20,inner sep=.3cm},
  to/.style={
    ->,>=stealth',shorten >=1pt,semithick,font=\sffamily\footnotesize
  },
  from/.style={
    <-,>=stealth',shorten >=1pt,semithick,font=\sffamily\footnotesize
  },
  tofrom/.style={
    <->,>=stealth',shorten >=1pt,semithick,font=\sffamily\footnotesize
  },
  every node/.style={align=center},
  squig/.style={->,line join=round,decorate, decoration={zigzag,
    segment length=8,amplitude=2,post=lineto,post length=2pt}}
}

\begin{document}
%
\title{Autotuning GPU Kernels via\\ Static and Predictive Analysis}

\author{\IEEEauthorblockN{Robert V. Lim,
Boyana Norris, and
Allen D. Malony}
\IEEEauthorblockA{Computer and Information Science\\
University of Oregon\\
Eugene, OR, USA\\
\{roblim1, norris, malony\}@cs.uoregon.edu}}


%


\maketitle

\begin{abstract}
Optimizing the performance of GPU kernels is challenging for both human programmers and code generators. For example, CUDA programmers must set thread and block parameters for a kernel, but might not have the intuition to make a good choice. Similarly, compilers can generate working code, but may miss tuning opportunities by not targeting GPU models or performing code transformations. Although empirical autotuning addresses some of these challenges, it requires extensive experimentation and search for optimal code variants. This research presents an approach for tuning CUDA kernels based on static analysis that considers fine-grained code structure and the specific GPU architecture features. Notably, our approach does not require any program runs in order to discover near-optimal parameter settings. We demonstrate the applicability of our approach in enabling code autotuners such as Orio to produce competitive code variants comparable with empirical-based methods, without the high cost of experiments.
\end{abstract}


%
\IEEEpeerreviewmaketitle

\section{Introduction}
Heterogeneous computing poses several challenges to the application developer.  Identifying which parts of an application are parallelizable on a SIMD accelerator and writing efficient data parallel code are the most difficult tasks.  For instance, CUDA programmers must set block and thread sizes for application kernels, but might not have the intuition to make a good choice.  With NVIDIA GPUs, each streaming multiprocessor (SM) has a finite number of registers, limited shared memory, a maximum number of allowed active blocks, and a maximum number of allowed active threads.  Variation in block and thread sizes results in different utilization of these hardware resources.  A small block size may not provide enough warps for the scheduler for full GPU utilization, whereas a large block size may lead to more threads competing for registers and shared memory.

Writing kernel functions requires setting block and thread sizes, and the difficulty is in deciding which settings will yield the best performance.  One procedure entails testing the kernel with block sizes suggested by the CUDA Occupancy Calculator (OCC)~\cite{cudaocc}.  Although the OCC takes into account the compute capability (NVIDIA virtual architecture) when calculating block sizes and thread counts, inaccuracies may arise because variations in runtime behavior may not be considered, which can potentially result in suboptimal suggested hardware parameters.

How do variations in runtime behavior arise? Accelerator architectures specialize in executing SIMD in lock-step.  When branches occur, threads that do not satisfy branch conditions are masked out. If the kernel programmer is unaware of the code structure or the hardware underneath, it will be difficult for them to make an effective decision about thread and block parameters. 

CUDA developers face two main challenges, which we aim to alleviate with the approach described in this paper.
First, developers must correctly select runtime parameters as discussed above.  A developer or user may not have the expertise to decide on parameter settings that will deliver high performance.  In this case, one can seek guidance from an optimization advisor.  The advisor could consult a performance model based on static analysis of the kernel properties, or possibly use dynamic analysis to investigate alternative configurations.  A second concern is whether the kernel implementation is not optimized yet. In this case, advice on parameter settings could still be insufficient because what is really required is a transformation of the kernel code itself to improve performance. For both concerns, static and dynamic analysis techniques are applicable.  However, to address the second concern, an autotuning framework based on code transformation is required.

This work presents our static analyzer that can be used by developers, autotuners, and compilers for heterogeneous computing applications.  Unlike most existing analysis techniques, our approach does not require any program runs to discover optimal parameter settings.  The specific contributions described in this paper include:
\begin{itemize}
\item A static analyzer for CUDA programs.
\item Predictive modeling based on static data.
\item Example use cases of the new methodology in an autotuning context.
\end{itemize}

Section~\ref{sec:motivation} provides background information, while Section~\ref{sec:methodology} defines the methodology of our static analyzer tool. Experimental setup and analysis are elaborated in Section~\ref{sec:experiments}; related work is discussed in Section~\ref{sec:related}, and Sections~\ref{sec:conclusion} and~\ref{sec:future} present our conclusions and future work plans.

\section{Background}
\label{sec:motivation}

This section briefly discusses the background for our research contributions, including the CUDA programming model, performance measurement approaches, and autotuning.

\begin{figure}
\centering
\vspace{-.1in}
\includegraphics[height=2in]{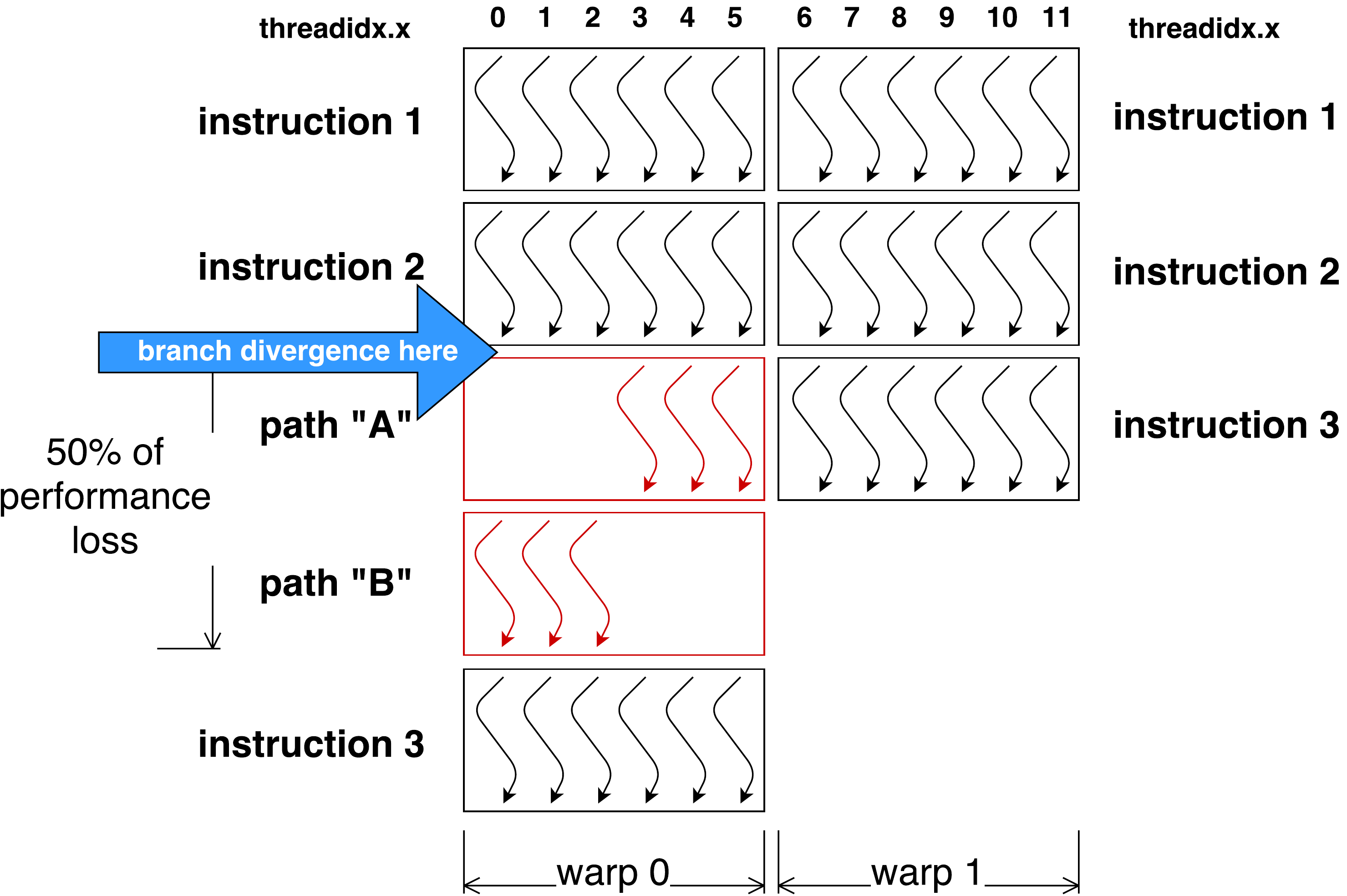}
\caption{
  Branch divergence problem and performance loss incurred.
}
\label{fig:diverg}
\end{figure}

\subsection{CUDA Programming Model and Control Flow Divergence}


In CUDA kernels, threads are organized in groups called blocks, which consists of one or more warps (each of which has 32 threads).  Each block is assigned to one of the GPU's streaming multiprocessors, and each SM is composed of multiple streaming processors, or multiprocessors (MP) that execute individual threads in SIMD.
%
%

In a given execution cycle, a SM executes instructions from one of the thread block's warps, where threads within a warp are executed together. However, if threads within a warp take different paths on conditional branches, execution of those paths become serialized.  In the worst case, only 1 of the 32 threads within a warp will make progress in a cycle.  Figure~\ref{fig:diverg} shows how performance is affected when branches diverge.  Measuring the occupancy of a kernel execution can determine whether branch divergence exists and suggest parameter adjustments to the program, a subject of this current work.

\subsection{GPU performance tools}
To date, GPU performance tools have mainly focused on the measurement and analysis of kernel execution, reporting time and counters associated with kernel execution.  For instance, the TAU Performance System provides scalable, profile and trace measurement and analysis for high-performance parallel applications~\cite{shende2006tau}, including support for CUDA and OpenCL codes~\cite{malony2011parallel}.  Even though profile measurements can help answer certain types of questions (e.g., how long did $foo()$ take?), improving performance requires more detailed information about the program structure.

While TAU and other profiling tools provide performance measurement \cite{ddt,nvprof,adhianto2010hpctoolkit}, they do not shed much light on the divergent branch behavior and its effects on making good decisions about thread and block sizes.  Our work introduces several static analysis techniques that delivers fine-grained information that can be used for predictive modeling.  These techniques include the ability to analyze instruction mixes and occupancy for estimating thread and register settings.  In a complementary approach (not discussed in this paper), we have also developed dynamic analysis techniques to compute instruction execution frequencies and control flow information \cite{lim2016tuning}.

In the remainder of this section, we discuss how we model different performance-relevant metrics by using primarily static analysis of CUDA binaries.

%
%

\begin{figure*}[thb]
\centering
\vspace{-.1in}
\includegraphics[height=2.5in]{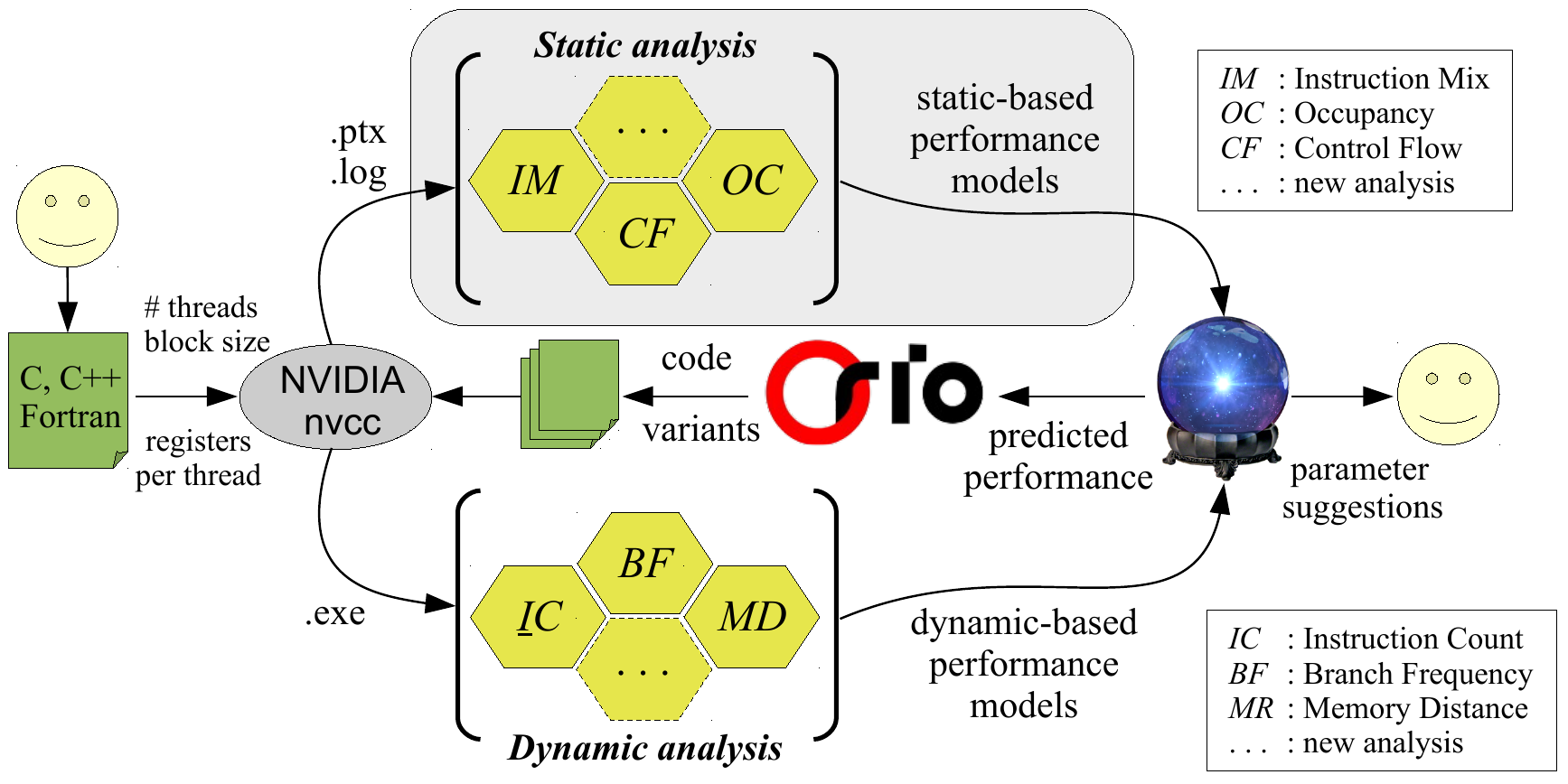}
\caption{
  Optimization framework for GPU kernels incorporating static and dynamic
  analysis, with autotuning and code transformation.
}
\label{figure:framework}
\end{figure*}

\subsection{Autotuning}

By themselves, performance models can produce adequate predictions of parameter settings, but can not change the kernel to improve performance.  Autotuning systems have been important in exploring alternative parameter choices by providing a kernel experimentation and optimization framework.  For example, the Orio autotuning framework~\cite{hartono2009annotation} generates many code variants for each kernel computation.  The objective of the GPU portions of Orio is to accelerate loops~\cite{chaimov2014toward,mametjanov2012autotuning} since loops consume a large portion of program execution time.  We use the term kernels to refer to deeply nested loops that arise frequently in a number of scientific application codes.  Existing C loops are annotated with transformation and tuning specifications.  Transformations are parameterized with respect to various performance constraints, such as block sizes, thread counts, preferred L1 sizes and loop unroll factors.  Each kernel specification generates a family of variant translations for each parameter and each variant is measured for its overall execution time, with the fastest chosen as the top performing autotuned translation.

The main challenge in the optimization space search is the costly empirical measurement of many code variants in autotuning systems.  The main contribution of our work is to demonstrate the use of static predictive models in autotuning, reducing the need for experimental performance testing.  

\section{Methodology}
\label{sec:methodology}

Figure~\ref{figure:framework} is a high-level depiction of our framework, which illustrates not only different processes involved, but also the analysis support and tradeoffs inherent in them.  For instance, providing a user with runtime parameters for kernel launch could engage static and/or dynamic analysis, but not necessarily code transformation.  Dynamic analysis would be expected to be more costly because empirical experiments would be involved. Transforming the implementation allows new variants to be explored, but these could be analyzed either statically or dynamically, or both.  However, it is in the integration of these models with an autotuning system that can transform the kernel code where the potential power for delivering optimizations is found.  

\medskip
\noindent{\bf Static Analysis}

Our static analysis approach consists of the following steps:

\begin{enumerate}\label{static_proc}
\item Extract kernel compilation information with \texttt{nvcc}'s \texttt{--ptxas-options=-v} flag.
\item Disassemble CUDA binary with \texttt{nvdisasm} for instruction operations executed.
\end{enumerate}

The subsequent sections define metrics resulting from our static analysis approach.

\subsection{Occupancy}
Threads, registers and shared memory are factors that influence a CUDA kernel's ability to achieve high occupancy.  In this section, we will group threads, warps, and blocks into one category for simplifying the discussion, although each term has its own distinct meaning.  Threads (\textit{T}) are the work units performing the computation, whereas warps (\textit{W}) are the schedulable units for the streaming multiprocessor and blocks (\textit{B}) consist of groups of warps.  Each has memory local to its level.  For instance, threads access private registers (\textit{R}), warps and blocks use shared memory (\textit{S}), and grids utilize global memory.  

The following subsections define factors that contribute to a kernel's GPU occupancy.   Table~\ref{tab:gpu} lists the GPUs used in this research, along with hardware features and associated notation.  We adopt the naming convention where superscripts denote the source of the variable, with subscripts as constraints of the variable.  Compute capability ($\mathit{cc}$) represents the GPU architecture family (also listed in Tab.~\ref{tab:gpu}), meaning \texttt{nvcc} will target an architecture based on the assigned compute capability flag (e.g. \texttt{-arch=sm\_xx}).  User input ($\mathit{u}$) includes threads, registers and shared memory parameters at compile time.  Active ($\mathit{*}$) represents the results provided by our static analyzer tool.  Occupancy is the metric we are calculating and is defined in the next subsections.

%
%
%
%

\subsubsection{Occupancy calculation}
The objective of the occupancy calculation is to minimize the number of active thread blocks per multiprocessor constrained by hardware resource $\mathit{\psi}$:


\begin{equation} \label{eq:bl_act}
B_{mp}^{\ast} = \min \left \{ \mathcal{G}_{\psi}(u) \right \},
\end{equation}

where $\mathcal{G}(\cdot)$ calculates the maximum allocable blocks for each SM, and $\psi= \{\psi_{W}, \psi_{R}, \psi_{S}\}$ denotes warps, registers, and shared memory.  Each $\mathcal{G}_\psi$ will be defined in Eqs.~\ref{eq:warps_sm},~\ref{eq:regs_sm}, and~\ref{eq:shmem_sm}.

\subsubsection{Definition of occupancy}
Occupancy is defined as the ratio of active warps on a SM to the maximum number of active warps supported for each SM:

\begin{alignat}{3}
\mathit{occ}_{mp} &=\frac{W_{mp}^{\ast}}{W_{mp}^{cc}}
\label{eq:occupancy}
\end{alignat}

where $W_{mp}^{\ast} = B_{mp}^{\ast} \times W_{B}$, with $B_{mp}^{\ast}$ as defined in Eq.~\ref{eq:bl_act} and $W_B=32$ for all GPUs (Tab.~\ref{tab:gpu}).  Note that in an ideal world, $\mathit{occ_{mp}}=1$.  However, in practice, occupancy rates are on average at 65-75\%, and should not be used in isolation for setting CUDA parameters \cite{volkov2010better}.  Occupancy is one of several metrics we incorporated in our static analyzer.


\subsubsection{Theoretical occupancy}
The number of blocks which can execute concurrently on an SM is limited by either warps, registers, or shared memory.

\paragraph{Warps per SM}

The SM has a maximum number of warps that can be active at once.  
To calculate the maximum number of blocks constrained by warps $\mathcal{G}_{\psi_{W}}$, find the minimum of blocks supported per multiprocessor and the rate of warps per SM and warps per block:

\begin{align}
\label{eq:warps_sm}
\mathcal{G}_{\psi_{W}}(T^{u}) &= \min\left \{ \mathit{B_{mp}^{cc}}, \left \lfloor{\frac{W_{sm}}{W_{B}}}\right \rfloor \right \}
\end{align}

where $W_{sm} = W_{mp}^{cc}$ and $W_B = \left \lceil \dfrac{T^{u}}{T_{W}^{cc}} \right \rceil$, with variables as defined in Table~\ref{tab:gpu}.


%
%
%

\paragraph{Registers per SM}

The SM has a set of registers shared by all active threads.  Deciding whether registers is limiting occupancy $\mathcal{G}_{\psi_{R}}$ is described by the following cases:

\begin{align}
\label{eq:regs_sm}
\mathcal{G}_{\psi_{R}}(R^{u}) &= 
\begin{cases}
0 &\text{if } R^{u} > R_{W}^{cc},\\
\left \lceil{\frac{R_{sm}}{R_B}}\right \rceil \times \left \lceil{\frac{R_{fs}^{cc}}{R_B^{cc}}}\right \rceil &\text{if } R^{u} > 0,\\
B_{mp}^{cc}  &\text{otherwise}.
\end{cases}
\end{align}

where $R_{sm} =  \left \lfloor{\dfrac{R_B^{cc}}{ \lceil R^{u} \times T_W^{cc} \rceil} } \right \rfloor$ and $R_B = \left \lceil \dfrac{T^{u}}{T_W^{cc}} \right \rceil$.  Case 1 represents when the user declares a register value beyond the maximum allowable per thread that is supported for the \textit{cc}, an illegal operation.  Case 2 describes when the user provides a valid register value, where we take the product of the number of registers per SM supported over the number of registers per block and the register file size per MP over the maximum register block supported in this architecture.  Case 3 is when the user does not provide a value, where the value is set to the thread block per multiprocessor supported by the \textit{cc}.


\begin{table} [t]
    \caption{GPUs used in this experiment.}
  \centering
\begin{tabular}{| >{\centering\arraybackslash} m{18pt} | >{\centering\arraybackslash} m{70pt}|>{\centering\arraybackslash}m{22pt} >{\centering\arraybackslash}m{22pt} >{\centering\arraybackslash}m{22pt} >{\centering\arraybackslash}m{22pt}|}\hline
\rowcolor{black!30}\Centering\bfseries Sym
  & \Centering\bfseries Parameter & \Centering\bfseries M2050 & \Centering\bfseries K20 & \Centering\bfseries M40 & \Centering\bfseries P100\\\hline
\ \textit{cc} & CUDA capability & 2 & 3.5 & 5.2 & 6.0  \\ 
\ & Global mem (MB)  & 3072 & 11520 & 12288 & 17066 \\ 
\ \textit{mp} & Multiprocessors & 14 & 13 & 24 & 56 \\ 
\ & CUDA cores / mp & 32 & 192 & 128 & 64 \\ 
\ & CUDA cores & 448 & 2496 & 3072 & 3584 \\ 
\ & GPU clock (MHz) & 1147 & 824 & 1140 & 405 \\ 
\ & Mem clock (MHz) & 1546 & 2505 & 5000 & 715 \\ 
& L2 cache (MB) & 0.786 & 1.572 & 3.146 & 4.194 \\ 
& Constant mem (B) & 65536 & 65536 & 65536 & 65536 \\ 
$S_B^{cc}$ & Sh mem block (B) & 49152 & 49152 & 49152 & 49152 \\ 
$R_{fs}^{cc}$ & Regs per block & 32768 & 65536 & 65536 & 65536 \\ 
$W_B$ & Warp size & 32 & 32 & 32 & 32 \\ 
$T_{mp}^{cc}$ & Threads per mp & 1536 & 2048 & 2048 & 2048 \\ 
$T_{B}^{cc}$ & Threads per block & 1024 & 1024 & 1024 & 1024 \\ 
$B_{mp}^{cc}$ & Thread blocks / mp & 8 & 16 & 32 & 32 \\ 
$T_{W}^{cc}$ & Threads per warp & 32 & 32 & 32 & 32 \\
$W_{mp}^{cc}$ & Warps per mp & 48 & 64 & 64 & 64 \\
$R_B^{cc}$ & Reg alloc size & 64 & 256 & 256 & 256 \\
$R_T^{cc}$ & Regs per thread & 63 & 255 & 255 & 255 \\   
\hline \hline
 & Family & Fermi & Kepler & Maxwell & Pascal \\   
\hline
\end{tabular}
  \label{tab:gpu}
\end{table}


\paragraph{Shared memory per SM}
Shared memory per thread is defined as the sum of static shared memory, the total size needed for all \texttt{\_\_shared\_\_} variables and dynamic shared memory.  If active blocks are constrained by shared memory, reducing \textit{S} per \textit{T} could increase occupancy.  To compute $\mathcal{G}_{\psi_S}$, take the ceiling of the shared memory per multiprocessor provided by its compute capability over the shared memory per block.  

\begin{align}
\label{eq:shmem_sm}
\mathcal{G}_{\psi_{S}}(S^{u}) &= 
\begin{cases}
0 &\text{if } S^{u} > S_B^{cc},\\
\left \lceil{\frac{S_{mp}^{cc}}{S_B}}\right \rceil &\text{if } S^{u} > 0,\\
B_{mp}^{cc}  &\text{otherwise}.
\end{cases}
\end{align}
where shared memory per block $S_B = \left \lfloor{S^{u}}\right \rfloor$, shared memory per SM $\mathit{S_{sm}} = S_{B}^{cc}$, and with cases following similarly to Eq.~\ref{eq:regs_sm}. 

Next, we describe instruction types and pipeline utilization and how increasing the number of active warps may adversely affect performance, since additional warps will oversubscribe the pipeline and force additional stalls.

\subsection{Instruction Mixes}

The instruction throughput on a GPU varies depending on its type, with memory operations typically achieving 32 instructions-per-cycle (IPC) and floating-point operations capable of 192 IPC.  Table~\ref{tab:ipc} lists IPCs, based on instruction type and compute capability \cite{geforce}. By definition, instruction throughput is the number of operations each SM can process per cycle.  In other words, an operation with a high throughput would cost less to issue than an operation with a lower instruction throughput.  In this work, we assign weights to instruction types based on its instruction throughput, defined as the reciprocal of IPC, or cycles-per-instruction (CPI).

\subsubsection{Instruction mix metrics}
Instruction mix is defined as the number of specific operations that a processor executes. Instruction mix-based characterizations have been used in a variety of contexts, including to select loop unrolling factors \cite{monsifrot2002machine,stephenson2005predicting}, unlike hardware counters which are prone to miscounting events \cite{lim2014computationally}.  In this work, we use instruction mixes to characterize whether a kernel is memory-bound, compute-bound, or relatively balanced.  Refer to \cite{lim2015identifying} for definitions for $\mathbf{O}_{fl}, \mathbf{O}_{mem}, \mathbf{O}_{ctrl}$, and $\mathbf{O}_{reg}$ according to category type.


The intensity (magnitude) of a particular metric can suggest optimal block and thread sizes for a kernel.  Memory-intensive kernels require a high number of registers, where a large block size consists of more registers per block.  The tradeoff with big block sizes is that fewer blocks can be scheduled on the SM.  Small block sizes will constrain the number of blocks running on the SM by the physical limit of blocks allowed per SM.  Compute-intensive kernels perform well with larger block sizes because the threads will be using GPU cores with fewer memory latencies.  Small block sizes will result in many active blocks running on the SM in a time-shared manner, where unnecessary switching of blocks may degrade performance.  For control-related synchronization barriers, smaller block sizes are preferred because many active blocks can run simultaneously on the SM to hide memory latency.  

\begin{table} [t]
    \caption{Instruction throughput per number of cycles.}
  \centering
\begin{tabular}{| >{\centering\arraybackslash} m{60pt} | >{\centering\arraybackslash} m{30pt}|>{\centering\arraybackslash}m{20pt} >{\centering\arraybackslash}m{20pt} >{\centering\arraybackslash}m{20pt} >{\centering\arraybackslash}m{20pt}|}\hline
\rowcolor{black!30}\Centering\bfseries Category
  & \Centering\bfseries Op & \Centering\bfseries SM20 & \Centering\bfseries SM35 & \Centering\bfseries SM52 & \Centering\bfseries SM60\\\hline
 FPIns32 & FLOPS &  32 & 192 & 128 & 64\\
FPIns64 & FLOPS & 16 & 64 & 4  & 32\\
CompMinMax & FLOPS & 32 & 160 & 64 & 32 \\
Shift, Extract, Shuffle, SumAbsDiff & FLOPS & 16 & 32 & 64 & 32 \\
Conv64 & FLOPS & 16 & 8 & 4 & 16 \\
Conv32 & FLOPS & 16 & 128 & 32 & 16 \\
LogSinCos & FLOPS & 4 & 32 & 32 & 16 \\
IntAdd32 & FLOPS & 32 & 160 & 64 & 32 \\
	\hline
TexIns, LdStIns, SurfIns & MEM & 16 & 32 & 64 & 16 \\
	\hline
PredIns, CtrlIns & CTRL & 16 & 32 & 64 & 16 \\
MoveIns & CTRL & 32 & 32 & 32 & 32 \\ 
\hline	
Regs & REG & 16 & 32 & 32  & 16 \\ 
\hline
\end{tabular}
  \label{tab:ipc}
\end{table}


\subsubsection{Pipeline utilization}
Each streaming multiprocessor (SM) consists of numerous hardware units that are specialized in performing a specific task.  At the chip level, those units provide execution pipelines to which the warp schedulers dispatch instructions.  For example, texture units provide the ability to execute texture fetches and perform texture filtering, whereas load/store units fetch and save data to memory.  Understanding the utilization of pipelines and its relation to peak performance on target devices helps identify performance bottlenecks in terms of oversubscription of pipelines based on instruction type.

The NVIDIA Kepler GK100 report \cite{geforce} lists instruction operations and corresponding pipeline throughputs per cycle.  Pipeline utilization describes observed utilization for each pipeline at runtime.  High pipeline utilization would indicate that the corresponding compute resources were used heavily and kept busy often during the execution of the kernel.

\subsubsection{Infer Kernel Execution Time}
Because the majority of CUDA applications are accelerated loops, we hypothesize that the execution time of a CUDA program is proportional to the input problem size $N$. Hence, 
\begin{alignat}{3}
\label{eq:timeexec}
\mathit{f(N)} &=\mathit{c}_{f}\cdot \mathit{\mathbf{O}_{fl}} + \mathit{c}_{m}\cdot \mathit{\mathbf{O}_{mem}} + \mathit{c}_{b}\cdot \mathit{\mathbf{O}_{ctrl}} + \mathit{c}_{r}\cdot \mathit{\mathbf{O}_{reg}}
\end{alignat}

where $\mathit{c}_{f}$, $\mathit{c}_{m}$, $\mathit{c}_{b}$, and $\mathit{c}_{r}$ are coefficients that represent the reciprocal of number of instructions that can execute in a cycle, or CPI.  Equation~\ref{eq:timeexec} represents how a program will perform for input size $N$ without running the application.

\medskip
\noindent{\bf Autotuning Integration}

\subsection{Static Search Analysis}
The static analyzer tool has been integrated in Orio to enable model-based pruning of the search space.  Current search algorithms in Orio include exhaustive, random, simulated annealing, genetic, and Nelder-Mead simplex methods~\cite{Norris:2007,hartono2009annotation}.  Adding this tool as a new search module in Orio demonstrates that our approach can easily be integrated into a general autotuning framework. 

We describe the overall search workflow of the static analyzer option.  Orio collects instruction counts for the CUDA kernel and computes the instruction mix metrics and occupancy rates, as defined in Sec.~\ref{static_proc}.  A rule-based model is invoked, which produces suggested parameter coordinates for Orio to search.  The heuristic that we employed was based on the kernel's computational intensity derived from collecting instruction mixes and its occupancy (Eq.~\ref{eq:bl_act}). Through empirical observation (discussed in Section~\ref{sec:discussion}), we have concluded that a threshold of $intensity > 4.0$ would benefit from upper ranges of thread values suggested by our static analyzer, whereas $intensity \leq 4.0$ would benefit from lower ranges of suggested thread values.  
An example search space is displayed in Figure~\ref{fig:perf_tune_orio}.  Without empirical testing, exhaustive search would be required to converge to optimal parameter settings, leading to a combinatorial explosive problem.  The goal is to minimize the range of parameter settings to tighten the search bounds, and our static analyzer feeds the necessary information to intelligently guide the autotuner.

\begin{table}[thb]
    \caption{A subset of features used for thread block classification.}
  \centering
  \begin{tabular}{|l|l|}
	\hline
	\multicolumn{1}{|p{0.15\columnwidth}|}{\centering\tabhead{Feature}} &
	\multicolumn{1}{p{0.15\columnwidth}|}{\centering\tabhead{Size}}\\
	\hline\hline
	\text{Thread Count} & 32 -- 1024 (with 32 increments) \\
	\text{Block size} \tablefootnote{Block sizes are compute capability specific.} & 24 -- 192 (with 16 increments) \\
	\text{Unroll loop factor} & \{1 -- 6\} \\
	\text{Compiler flags} & \{\texttt{`'}, \texttt{`-use\_fast\_math'}\} \\
	\hline\hline	
	\text{Instructions} & \{FLOPS, memory, control\} \\
	\text{Occupancy calculation} & \{registers, threads, OCC rate, etc.\} \\

	\hline
	
  \end{tabular}

  \label{tab:features}
\end{table}

\begin{figure}
  \centering \small
\begin{verbatim}
/*@ begin PerfTuning (
  def performance_params {
    param TC[]  = range(32,1025,32);
    param BC[]  = range(24,193,24);
    param UIF[] = range(1,6);
    param PL[]  = [16,48];
    param SC[]  = range(1,6);
    param CFLAGS[] = ['', '-use_fast_math'];
  }
  ...
) @*/
\end{verbatim}
  \caption{Performance tuning specification in Orio.}
  \label{fig:perf_tune_orio}
\end{figure}


\section{Experimental Setup and Analysis}
This section reports on the autotuning execution environment for the CUDA kernels listed in Table~\ref{tab:kernels}.  Results comparing our static analyzer approach with the existing methods are also reported.

\label{sec:experiments}

\subsection{Environment}
Orio was used to generate and autotune CUDA implementations by varying the feature space listed in Table ~\ref{tab:features}.  The details of CUDA code generation and autotuning with Orio are described in~\cite{mametjanov2012autotuning}.  The $\mathit{TC}$ parameter specifies the number of simultaneously executing threads. $\mathit{BC}$ is the number of thread blocks (independently executing groups of threads that can be scheduled in any order across any number of SMs) and is hardware-specific. $\mathit{UIF}$ specifies how many times loops should be unrolled; $\mathit{PL}$ is the L1 cache size preference in KB. 
%

For each code variant, the experiment was repeated ten times, and the fifth overall trial time was selected to be displayed.  The execution times were sorted in ascending order and the ranks were split along the 50th percentile.  Rank 1 represents the upper-half of the 50th percentile (good performers), while Rank 2 represents the lower portion (poor performers).  On average, the combination of parameter settings generated 5,120 code variants.  The GPUs used in this work are listed in Table~\ref{tab:gpu} and include the Fermi M2050, Kepler K20, Maxwell M40, and Pascal P100.  Subsequently we will refer to the GPUs by the architecture family name (Fermi, Kepler Maxwell, Pascal).  CUDA \texttt{nvcc} v7.0.28 was used as the compiler.  Each of the benchmarks executed with five different input sizes, where all benchmarks consisted of inputs \{32, 64, 128, 256, 512\}, except ex14FJ, which had inputs \{8, 16, 32, 64, 128\}.

To demonstrate our approach, we considered the kernels described in Table~\ref{tab:kernels}. Because the chosen kernels (except ex14FJ, which is application-specific) contribute significantly to the overall execution time of many different applications, tuning 
these kernels can result in significant overall application performance improvements.

\begin{table*}
   \caption{Kernel specifications.}
  \label{tab:kernels}

  \centering
  \begin{tabular}{|c|l|l|c|c|c|c|}
        \hline
        \multicolumn{1}{|p{0.15\columnwidth}|}{\centering\tabhead{Kernel}} &
        \multicolumn{1}{p{0.15\columnwidth}|}{\centering\tabhead{Category}} &
        \multicolumn{1}{p{0.15\columnwidth}|}{\centering\tabhead{Description}} &
        \multicolumn{1}{p{0.15\columnwidth}|}{\centering\tabhead{Operation}}\\
        \hline\hline
        \texttt{atax} & Elementary linear algebra & Matrix transpose, vector multiplication & $y=A^{T}(Ax)$ \\
        \texttt{BiCG} & Linear solvers & Subkernel of BiCGStab linear solver & \pbox{20cm}{$q=Ap,$\\ $s=A^{T}r$} \\
        \texttt{ex14FJ} & 3-D Jacobi computation & Stencil code kernels & \pbox{20cm}{$F(x)=A(x)x-b=0,$\\ $A(u)v \simeq -\bigtriangledown (\kappa(u)\triangledown v)$} \\
       

        \texttt{MatVec2D} & Elementary linear algebra & Matrix vector multiplication & $y=Ax$ \\
        
        \hline
  \end{tabular}
\end{table*}

\begin{figure*}
\centering
\vspace{-.1in}
\includegraphics[height=2.5in,width=7.5in]{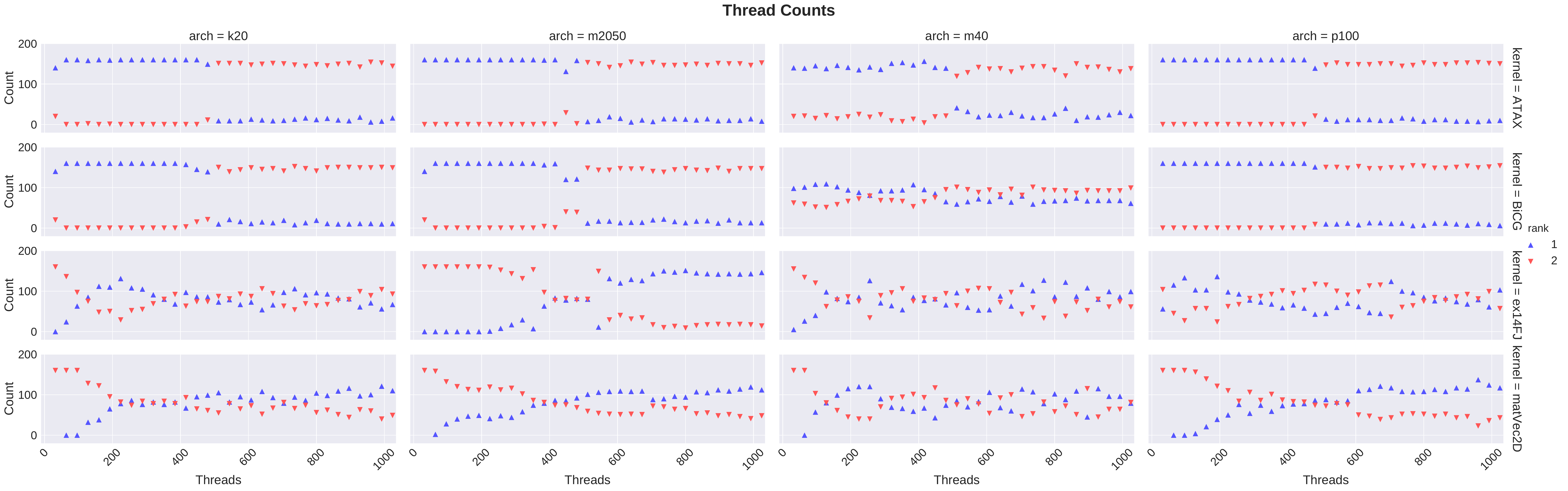}
\caption{
Thread counts for Orio autotuning exhaustive search, comparing architectures and kernels.
}
\label{fig:tc}
\end{figure*}

\begin{figure}
\centering
\vspace{-.1in}
\includegraphics[height=3in]{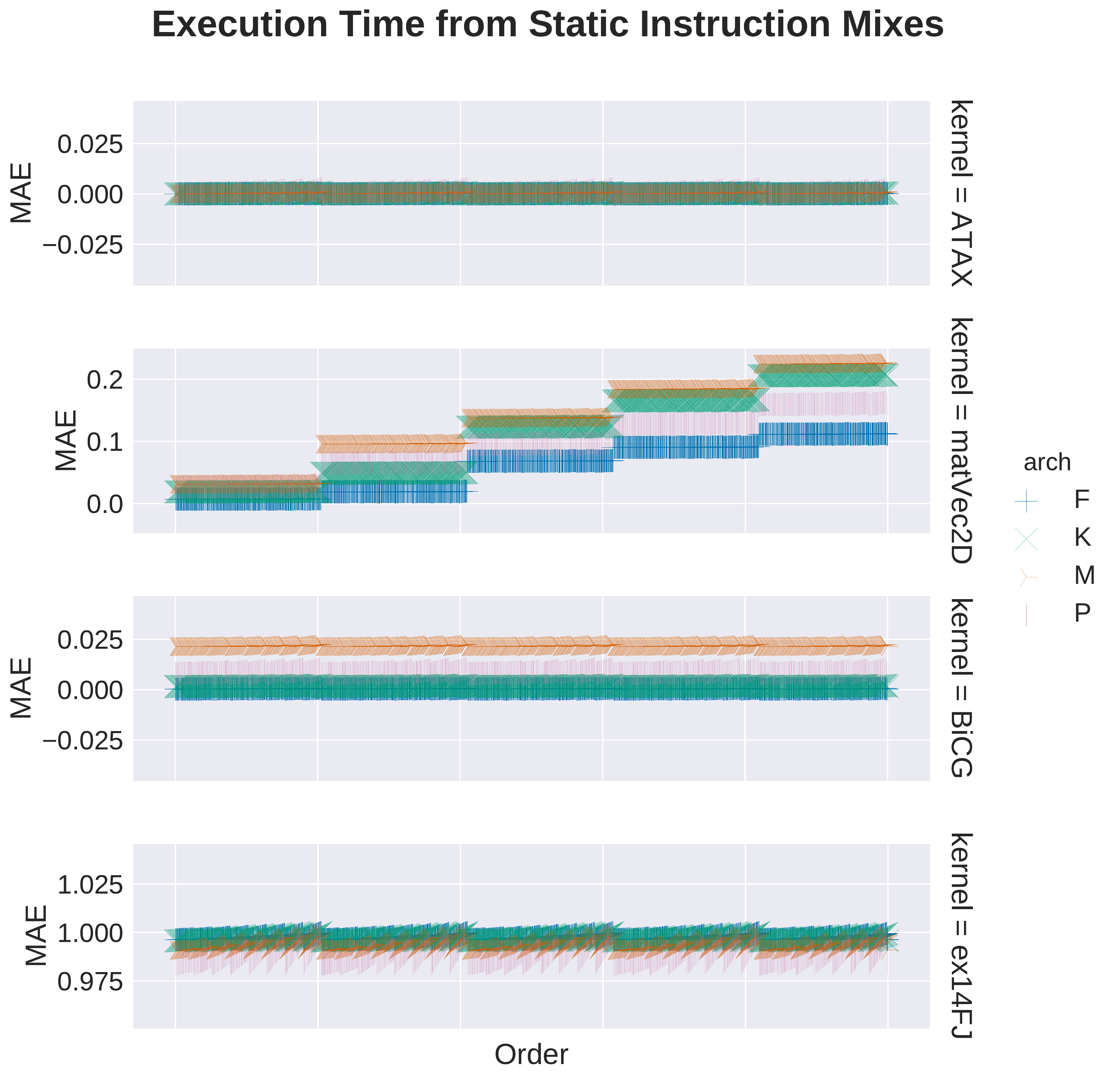}
\caption{
Time-to-instruction mix ratio, comparing architectures and kernels.
}
\label{fig:time_imix}
\end{figure}

\begin{table*}
  \centering
    \caption{Statistics for autotuned kernels for top performers (top half) and poor performers (bottom half), comparing GPU architecture generations.}
  \label{tab:breakdown}
  \begin{tabular}{|r|r|rrr|rrr|rrr|}
    \hline
    \multicolumn{2}|c&\multicolumn{3}{c|}{Occupancy}&\multicolumn{3}{c|}{Register Instructions}&\multicolumn{3}{c|}{Threads}\\
    \arrayrulecolor{black}
    \multicolumn{2}|c& Mean & Std Dev & Mode & Mean & Std Dev & Allocated & 25th & 50th & 75th\\
    \hline \hline
     \multirow{2}{*}& Fer & 77.46 & 24.18 & 100.00 & 39613.1 & 35673.2 & 21 & 152 & 272 & 416\\
     {ATAX} & Kep & 85.21 & 19.03 & 93.75 & 34833.1 & 30954.5 & 27 & 160  & 288 & 416\\
     & Max & 90.59 & 11.87 & 93.75 & 104285.9 & 85207.1 & 30 & 160 & 320 & 448\\
     & Pas & 90.86 & 12.24 & 93.75 & 227899.7 & 202120.2 & 30 & 152 & 272 & 392\\     
     \hline
    \multirow{2}{*} & Fer & 60.55 & 15.54 & 75.00 & 35321.3 & 32136.6 & 27 & 160 & 288 & 416\\
     {BiCG} & Kep & 85.14 & 19.05 & 98.44 & 35485.7 & 31535.9 & 28 & 160 & 288 & 416\\
     & Max & 89.09 & 11.50 & 98.44 & 158963.8 & 135681.2 & 32 & 224 & 448 & 736\\     
     & Pas & 90.93 & 12.19 & 93.75 & 228350.6 & 201865.8 & 30 & 152 & 272 & 392\\     
     \hline
    \multirow{2}{*} & Fer & 53.69 & 8.83 & 62.50 & 98418.5 & 45166.64 & 30 & 608 & 768 & 896\\
     {ex14FJ} & Kep & 88.44 & 9.98 & 93.75 & 54345.4 & 47526.8 & 31 & 288 & 512 & 768\\
     & Max & 89.23 & 9.61 & 98.44 & 4141130.6 & 158537.4 & 28 & 320 & 608 & 832\\
     & Pas & 89.04 & 11.10 & 98.44 & 4335986.6 & 409162.6 & 32 & 192 & 480 & 768\\     
     \hline
    \multirow{2}{*} & Fer & 72.21 & 14.17 & 87.50 & 307425.50 & 69330.06 & 23 & 448 & 640 & 864\\
      {matVec2D} & Kep & 89.29 & 8.17 & 96.88 & 274359.93 & 65373.88  & 23 & 416 & 640 & 864\\
	  & Max & 89.53 & 9.22 & 98.43 & 693752.81 & 146799.80 & 18 & 288 & 576 & 800\\
	  & Pas & 88.42 & 9.08 & 90.63 & 1264815.81 & 316252.38 & 18 & 480 & 672 & 864\\	  
	   \hline
    \hline
     \multirow{2}{*} & Fer & 74.23 & 15.98 & 100.00 & 102946.9 & 58009.0 & 21 & 640 & 768 & 896\\
      {ATAX} & Kep & 86.27 & 10.97 & 93.75 & 89906.9 & 51102.5 & 27 & 640 & 768 & 896\\                                                
      & Max & 87.04 & 10.09 & 87.50 & 253714.1 & 151973.5 & 30 & 608 & 736 & 896\\
      & Pas & 86.77 & 9.54 & 87.50 & 605300.3 & 337615.5 & 30 & 640 & 768 & 896\\      
    \hline
    \multirow{2}{*}&Fer & 56.12 & 10.73 & 66.67 & 35321.3 & 32136.6 & 27 & 608 & 768 & 896\\
      {BiCG} & Kep & 86.34 & 10.93 & 93.75 & 89254.3 & 51141.2 & 28 & 608 & 768 & 896\\                                                
      & Max & 88.55 & 10.80 & 93.75 & 199036.3 & 149373.1 & 32 & 352 & 608 & 832\\
      & Pas & 86.70 & 9.57 & 87.5 & 605169.4 & 338092.4 & 30 & 640 & 768 & 896\\      
      \hline  
    \multirow{2}{*}&Fer & 55.55 & 14.03 & 62.50 & 26321.5 & 21137.2 & 30 & 152 & 288 & 448\\
     {ex14FJ} & Kep & 83.05 & 19.21 & 93.75 & 70394.6 & 51953.5 & 31 & 256 & 544 & 800\\
     & Max & 88.40 & 12.50 & 93.75 & 4079589.4 & 120401.0 & 28 & 224 & 480 & 704\\
     & Pas & 88.59 & 11.22 & 93.75 & 4359934.4 & 241618.2 & 32 & 352 & 544 & 800\\     
    \hline
    \multirow{2}{*}&Fer & 68.93 & 21.93 & 87.50 & 210334.50 & 47850.90 & 23 & 160 & 352 & 672\\
      {matVec2D} & Kep & 82.19 & 19.54 & 93.75 & 219636.56 & 57185.33 & 23 & 160 & 384 & 704\\                                                
      & Max & 88.09 & 12.77 & 93.75 & 645687.18 & 137182.93 & 18 & 224 & 480 & 736\\
      & Pas & 89.22 & 12.89 & 93.75 & 877505.0 & 225900.05 & 18 & 160 & 320 & 576\\
       \hline
  \end{tabular}
\end{table*}

\begin{table}
  \centering
    \caption{Error rates when estimating dynamic instruction mixes from static mixes.}
  \begin{tabular}{|c|c|c|c|c|c|c|c|}
    \hline
    \multicolumn{2}|c&\multicolumn{4}{c|}{Metrics}\\
    \arrayrulecolor{black}
    \cline{3-6}
    \multicolumn{2}|c&FLOPS&MEM&CTRL&Itns\\
    \hline
    \multirow{2}{*}{ATAX}&Fer &0.07&1.69&2.01&3.4\\
    \cline{2-6}
                         &Kep &0.11&1.75&2.20& 3.4\\
    \cline{2-6}
                         &Max &0.23&0.06&0.12& 1.8 \\
                             \hline
    \multirow{2}{*}{BiCG}&Fer &0.03&3.68&2.40& 1.8 \\
    \cline{2-6}
                         &Kep &0.02&3.80&2.67& 1.8 \\
	\cline{2-6}
                         &Max &0.57&1.30&0.06& 1.3 \\
    \hline
    \multirow{2}{*}{ex14FJ}&Fer &0.20&0.14&0.00& 12.7\\
    \cline{2-6}
                         &Kep &1.01&0.18&0.21& 12.7\\
    \cline{2-6}
                         &Max &1.97&0.14&0.89& 16.3\\                         
    \hline
    \multirow{2}{*}{matVec2D}&Fer &0.04&0.92&0.80& 4.6\\
    \cline{2-6}
                         &Kep &0.07&0.97&0.99& 4.6\\
    \cline{2-6}
                         &Max &0.29&0.06&0.36& 7.2\\                         
    \hline
  \end{tabular}
  \label{tab:error_rate_imix}
\end{table}

\subsection{Discussion}
\label{sec:discussion} 
We empirically autotuned the kernels listed in Table~\ref{tab:kernels} using exhaustive search and uncovered distinct ranges for block and thread sizes, based on ranking.  The dynamic analysis of autotuning is displayed in Figure~\ref{fig:tc}, projecting thread settings and frequency for each kernel, and comparing various architectures.  In general, ATAX and BiCG kernels performed well in lower thread range settings, whereas matVec2D performed better with higher thread settings.  The ex14FJ is a more complex kernel\footnote{The ex14FJ kernel is the Jacobian computation for a solid fuel ignition simulation in 3D rectangular domain.}, and thread behavior patterns for Rank 1 were less apparent.

Table~\ref{tab:breakdown} reports statistics on occupancy, registers, and threads for all benchmarks and architectures.  The top half represents good performers (Rank 1), whereas the bottom half represents poor performers (Rank 2).  In general, occupancy did not seem to matter much, since the reported means were somewhat similar for both ranks, with Fermi achieving low \textit{occ} for all kernels.  However, register instructions varied considerably, with Rank 1 consisting of lower mean and standard deviations, versus Rank 2 which had higher values.  Thread behavior patterns were apparent when comparing Rank 1 and Rank 2.  For instance, one could conclude that ATAX and BiCG prefers smaller range thread sizes, whereas ex14FJ prefers higher ranges.

Figure~\ref{fig:time_imix} illustrates the use of static instruction mixes to predict execution time.  Execution time was normalized and sorted in ascending order (x-axis).  The mean absolute error was used to estimate execution time based on static instruction mixes.  Equation~\ref{eq:timeexec} was used to calculate the instruction mix ratio, which consisted of weighting instructions according to its number of achievable executed instructions per clock cycle.  In general, our model was able to estimate the execution time within a reasonable margin of error, including \texttt{ex14FJ} with \textit{MAE} near $1.00$, which validates instruction mixes as good indicators of kernel execution performance.

Table~\ref{tab:error_rate_imix} reports the error rates calculated, using sum of squares, when estimating dynamic behavior of the kernel from static analysis of the instruction mix.   Intensity is also displayed in the last column and is defined as the ratio of floating-point operations to memory operations. Although our static estimator performed poorly for BiCG (memory, control ops), our static analysis, driven by Equation~\ref{eq:timeexec}, closely matches that of the observed dynamic behavior for the other kernels.

\subsection{Improved Autotuning with Static Analyzer}

\begin{table}
  \centering
    \caption{Suggested parameters to achieve theoretical occupancy.}
  \begin{tabular}{|c|c|c|c|c|c|c|}
    \hline
    \arrayrulecolor{black}
    \multicolumn{2}|& &$T^{*}$ & $[R^{u}:R^{*}]$ & $S^{*}$ &$occ^{*}$\\
    \hline
    \multirow{4}{*}{AT}&Fer&192, 256, 384, 512, 768 & [21 : 0] & 6144 &1\\
    \cline{2-6}
                         &Kep & 128, 256, 512, 1024 & [27 : 5] & 3072 & 1\\
    \cline{2-6}
                         &Max& 64, 128, 256, 512, 1024 & [30 : 2] & 1536 & 1\\ 
   \cline{2-6}                         
                         &Pas& 64, 128, 256, 512, 1024 & [30 : 2] & 1536 & 1\\                         
                             \hline
    \multirow{4}{*}{Bi}&Fer & 192, 256, 384, 512, 768 & [27 : 0] & 8192 & .75 \\
    \cline{2-6}
                         &Kep &128, 256, 512, 1024 & [28 : 4] & 3072 & 1\\
    \cline{2-6}
                         &Max &64, 128, 256, 512, 1024 & [32 : 0] & 12288 &.71\\
                             \cline{2-6}
                         &Pas &64, 128, 256, 512, 1024 & [30 : 2] & 1536 & 1\\
    \hline
    \multirow{4}{*}{ex}&Fer &192, 256, 384, 512, 768 & [30 : 0] & 24576 &.71\\
    \cline{2-6}
                         &Kep &128, 256, 512, 1024 & [31 : 1] & 3072 & 1\\
    \cline{2-6}
                         &Max &64, 128, 256, 512, 1024 & [28 : 4] & 1536 & 1\\
    \cline{2-6}
                         &Pas &64, 128, 256, 512, 1024 & [32 : 0] & 1536 & 1\\                         
    \hline
    \multirow{4}{*}{ma}&Fer &192, 256, 384, 512, 768 & [20 : 1] & 12288 & .92\\
    \cline{2-6}
                         &Kep &128, 256, 512, 1024 & [20 : 11] & 3072 & 1\\
    \cline{2-6}
                         &Max &64, 128, 256, 512, 1024 & [13 : 18] & 1536 & 1\\
    \cline{2-6}
                         &Pas &64, 128, 256, 512, 1024 & [15 : 17] & 1536 & 1\\                         
    \hline
  \end{tabular}
  \label{tab:metrics}
\end{table}

\begin{figure*}[htb]
	\centering
	\includegraphics[width=1.0\textwidth]{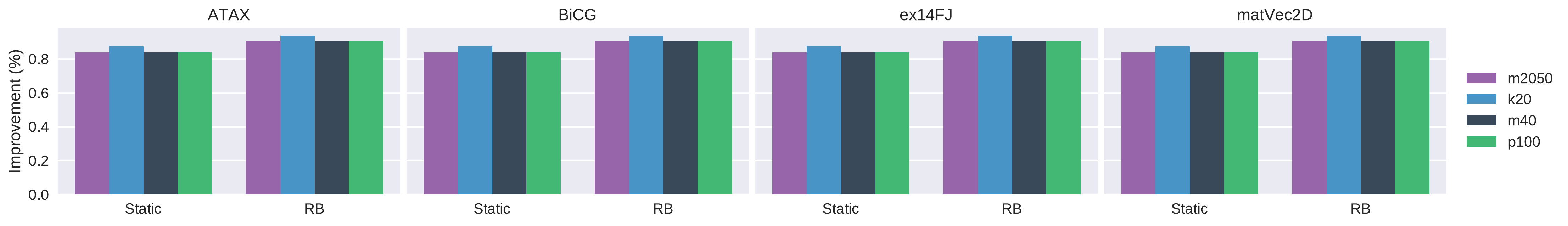}
	\centering
	\caption{Improved search time over exhaustive autotuning, comparing static and rule-based approaches.}
	\centering
	\label{fig:search}
\end{figure*}

Finally, we wanted to determine whether our static analyzer tool could be used to improve the efficiency and effectiveness of Orio.  We use the exhaustive empirical autotuning results from Sec.~\ref{sec:discussion} as the baseline for validating whether our search approach could find the optimal solution. 

Table~\ref{tab:metrics} reports static information for register usage and intensity for each kernel, as well as the thread parameters suggested by our static analyzer, comparing different architectures.  $T^{*}$ displays the suggested thread ranges for the kernel that would yield $occ^{*}$.  $[R^{u} : R^{*}]$ displays the number of registers used and its increase potential.  $S^{*}$ displays (in \textit{KB}) the amount of shared memory that could be increased to achieve theoretical occupancy.  

\begin{figure*}[htb]
	\centering
	\includegraphics[width=1.0\textwidth]{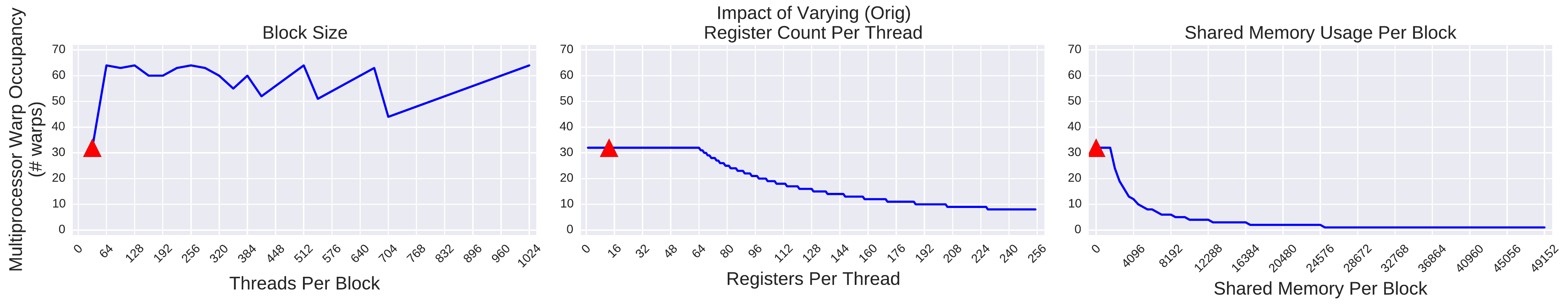}
	\includegraphics[width=1.0\textwidth]{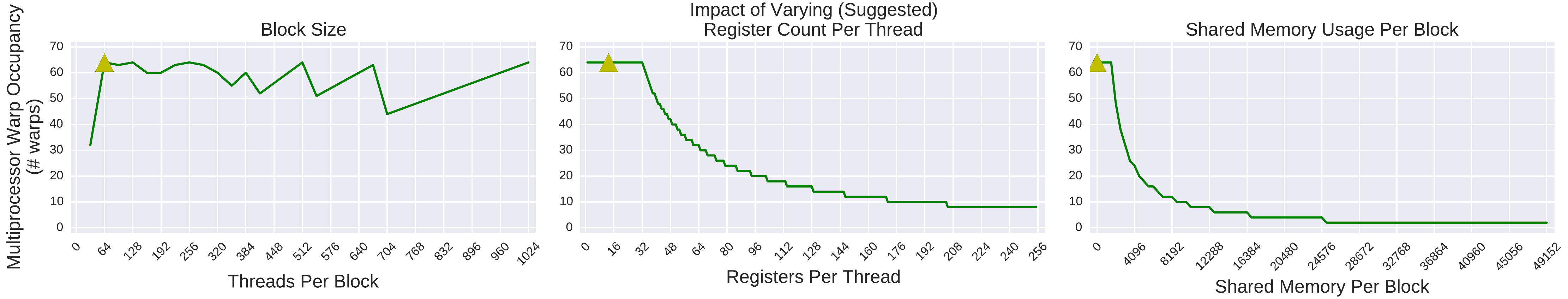}
	\centering
	\caption{Occupancy calculator displaying thread, register and shared memory impact for current (top) and potential (bottom) thread optimizations for the purposes of increasing occupancy.}
	\centering
	\label{fig:cuda_occ}
\end{figure*}

The basis of our contribution is that the instruction mix and occupancy metrics from our static analyzer gets fed into the autotuner.  In general, an exhaustive autotuning consists of $\prod_{i=1}^{m}\mathbf{X}_{i}$ trials, where $\mathbf{X}_i$ represents a parameter, each having $m$ options.  In the case of ATAX, five thread settings were suggested for Fermi and Maxwell, which represents a 84\% improvement, and Kepler representing a 87.5\% improvement, with the search space reduced from 5,120 to 640.  
The search space could be reduced further by invoking our rule-based heuristic.  Figure~\ref{fig:search} displays the overall results of the improved search module.  The first set displays how the static based method improves near 87.5\%.  When combining with the rule-based heuristic, the search space is further reduced, which results in a 93.8\% overall improvement.  Figure~\ref{fig:cuda_occ} displays the occupancy calculator for the ATAX kernel, comparing the current kernel and the potentially optimized version.  

The model-based search space reduction does involve generating and compiling the code versions, but it does \textbf{not} require executing them. Note that empirical testing typically involves multiple repeated executions of the same code version, hence the time saved over exhaustive search is approximately $t*r$, where $t$ is the average trial time and $r$ is the number of repetitions. 
Even when not using exhaustive search, our new technique can be used as the first stage of the regular empirical-based autotuning process to dramatically reduce the search space, significantly speeding up the entire process and increasing the likelihood of finding a global optimum. 
Unlike runtime measurement, which requires several runs of each test, static analysis does not suffer from the effects of noise and hence only has to be performed once on each code version. The search space reduced through static binary analysis can then be explored using one of the existing search methods. If it's feasible and desirable to determine the optimal value, then exhaustive search is appropriate, otherwise one of the other methods such as Nelder-Mead simplex or random can be used to strictly control the time spent autotuning.


\section{Related Work}
\label{sec:related}

Several prior efforts have attempted to discover optimal code forms and
runtime parameter settings for accelerator-based programming models, typically by taking a domain-specific approach.  
For instance, Nukada and
Matsuoka demonstrated automated tuning for a CUDA-based 3-D FFT library
based on selection of optimal number of threads~\cite{3Dfft}.  Tomov et
al. developed the MAGMA system for dense linear algebra solvers for GPU
architectures, which incorporates a DAG representation and empirical-based
search process for modeling and optimization~\cite{magma}.  The use of
autotuning systems based on program transformations, such as
Orio~\cite{hartono2009annotation} and CHiLL~\cite{chill}, enable optimization exploration on more general application code and across accelerator
architectures~\cite{chaimov2014toward}.  However, the complexity of the optimization
space and the cost of empirical search is high.  A recent work on autotuning GPU kernels focuses on loop scheduling and is based on the OpenUH compiler~\cite{xu2016analytical}.  Our approach attempts to
leverage more static code analysis to help better inform an autotuning
process, thereby reducing the dependence on pure dynamic measurement and
analysis to generate performance guidance.

The NVIDIA CUDA Toolkit~\cite{cuda-toolkit} includes occupancy calculation
functions in the runtime API that returns occupancy estimates for a given
kernel.  In addition, there are occupancy-based launch configuration
functions that can advise on grid and block sizes that are expected to
achieve the estimated maximum potential occupancy for the kernel.  Because
these functions take as input intended per-block dynamic shared memory usage
and maximum block size (in addition to knowing user-defined registers per
thread), it is possible to retrieve a set of configuration choices.  It is
important to note that the CUDA Occupancy Calculator/API takes into account
the GPU architecture being used.  Thus, we can integrate the estimates it generates over
the full range of options (e.g., letting registers per thread to be
variable) with the other static models.




%

A project closely related to ours is STATuner~\cite{statuner}, 
which identifies a feature set of static metrics that characterize a
CUDA kernel code and uses machine learning to build a classifier model
trained on a CUDA benchmark suite.  Kernel codes are compiled in LLVM and
static analysis of the generated binary code and IR provide metrics for
instruction mix, loops, register usage, shared memory per block, and thread
synchronization.  The classifier model inputs these metric features for a
new kernel to predict which block size would give the best performance.
STATuner is shown to give smaller average error compared to NVIDIA's CUDA
Occupancy Calculator/API.  Only a single block size is predicted by
STATuner, whereas the Occupancy Calculator/API offers block size choices
given user input about registers per thread and per-block shared memory.
Our approach differs in several respects.  First, static analysis is done
on the PTX code generated by the NVIDIA nvcc compiler, rather that on the
upper level source code (as seen in LLVM).  While there are some benefits 
in incorporating higher-level code information, nvcc produces
different PTX code for different GPU architectures, allowing
hardware-specific code effects to be seen.  Furthermore, our static
analysis extracts metrics similar to STATuner, but also builds a CFG to
help understand flow divergence \cite{lim2016tuning}.  Second, our prediction models are based
on estimating performance given the instruction mix, control flow, and
problem size.  They are not based on learned classifiers.  Third, the
objective of our work is to integrate predictive models in an autotuning
framework, beyond just giving a single block size result to the user.

Milepost GCC~\cite{fursin2011milepost} is a publicly-available open-source machine learning-based compiler for C (but not CUDA) that 
extracts program features and exchanges optimization data with the cTuning.org open public repository. It automatically adapts the 
internal optimization heuristic at function-level granularity to improve execution time, code size and compilation time of a new program 
on a given architecture.
 
 
The Oxbow toolkit~\cite{sreepathi2014application} is a collection of tools to empirically characterize (primarily CPU) application behaviors, including computation, communication, memory capacity and access patterns.
 The eventual goal is to build a repository that users can upload
and access their datasets, and can provide analysis, plots, suggested
parameters, etc.

\section{Conclusion}\label{sec:conclusion}
Getting the most performance out of applications is important for code generators and end users, but the process in making the best settings is often convoluted.  With our static analyzer tool, we show its accuracy in estimating the runtime behavior of a kernel without the high costs of running experiments.  Using our tool, we've identified the computational intensity of a kernel, constructed a control flow graph, estimated the occupancy of the multiprocessors, and suggested optimizations in terms of threads and register usage.  Finally, we've shown how the integration of our static analyzer in the Orio autotuning framework improved the performance in narrowing the search space for exploring parameter settings.

The field of heterogeneous accelerated computing is rapidly changing, and we expect several disruptions to take place  with the introduction of 3D-memory subsystems, point-to-point communication, and more registers per computational cores.  Traditional approaches to measuring performance may no longer be sufficient to understand the behavior of the underlying system.  Our static analyzer approach can facilitate optimizations in a variety of contexts through the automatic discovery of parameter settings that improve performance.

\section{Future Work}\label{sec:future}

The optimization spectrum is a continuum from purely static-based methods
to ones that incorporate empirical search across an optimization landscape.
In general, the objective of our work is on exploring the tradeoffs
involving optimization accuracy and cost over this spectrum, with a
specific focus on how well purely static methods perform as a guide for
autotuning.  While static analysis side-steps the need for empirical
testing, it is not to say that static models can not be informed by prior
benchmarking and knowledge discovery.  We will investigate several avenues
for enhancing our static models, including algorithm-specific optimizations
and machine learning for code classification.

Furthermore, we regard the methodology we have developed as a knowledge
discovery framework where the degree of empirical testing can be ``dialed
in'' during the autotuning process, depending on what the user accepts.  By
recording the decisions and code variants at each step, it is also possible
to replay tuning with empirical testing for purpose of 
validation.  In this way, the framework can continually evaluate the static
models and refine their predictive power.  We will further develop this capability.

While our static analysis tools will working with any CUDA kernel code, the
real power of our approach is in the ability to transform the code in Orio.
However, this requires the source to be in a particular input form.  We are
exploring source analysis technology \cite{CERE} to translate kernel code to the input required by Orio, thereby allowing any kernel to be a candidate
for CUDA autotuning.


\section*{Acknowledgment}
We want to thank the anonymous reviewers for the insightful comments.  We also want to thank NVIDIA for providing early access to CUDA 7.5 and to the PSG clusters.  This work is supported by American Society of Engineering Education (ASEE) and Department of Energy (Award \#DE-SC0005360) for the project ``Vancouver 2:  Improving Programmability of Contemporary Heterogeneous Architectures.''
%



%
%
%

\bibliographystyle{IEEEtran}
\bibliography{references}

\end{document}